\documentclass{epl}

\title{ Quantum phase transition in the dioptase magnetic lattice}
\author{ Claudius Gros\inst{1} \and Peter Lemmens\inst{2} \and 
         K.-Y. Choi\inst{3} \and G. G\"untherodt\inst{3} \and
	 M. Baenitz\inst{4} \and H.H. Otto\inst{5}
       }
\institute{
\inst{1} Fachbereich Physik, Postfach 151150, Universit\"at des Saarlandes,
       66041 Saarbr\"ucken, Germany \\
\inst{2} IMNF, TU Braunschweig, D-38106 Braunschweig, Germany\\
\inst{3} II. Physikalisches Institut, RWTH-Aachen, Templergraben 55,
               52056 Aachen, Germany\\
\inst{4} Max-Planck-Institut f\"{u}r Chemische Physik fester Stoffe,
	               MPI-CPfS, D-01187 Dresden, Germany\\
\inst{5} Institut für Mineralogie und Mineralische Rohstoffe, TU Clausthal,
                     D-38678 Clausthal-Zellerfeld, Germany
         }

\begin{document}

\maketitle

\begin{abstract}
The study of quantum phase transitions \cite{sac99}, which
are zero-temperature phase transitions between
distinct states of matter, is
of current interest in research since it allows for
a description of low-temperature properties based
on universal relations.
Here we show that the crystal
green dioptase Cu$_{6}$Si$_{6}$O$_{18}\cdot$6H$_2$O, known
to the ancient Roman as the gem of Venus,
has a magnetic crystal structure, formed by the
Cu(II) ions, which allows for a quantum phase transition
between an antiferromagnetically ordered state and a
quantum spin liquid.
\end{abstract}

The gem-stone dioptase 
Cu$_{6}$Si$_{6}$O$_{18}\cdot$6H$_2$O is a transparent
green mineral build up from Si$_6$O$_{18}$ single rings 
on a lattice, which sandwiches six-membered water rings
down the (crystallographic) $c$-direction
\cite{hei55,rib77,diop_crystal}.
The magnetic  Cu(II) ions are located between the
$\rm Si_6O_{18}$ rings and form chiral chains along $c$,
placed on an $ab$-honeycomb net and are there
edge-sharing connected forming Cu(II) dimers.

\begin{figure}[t]
\hbox to\hsize{%
\hfill
\rotatebox{0}{\resizebox{!}{0.65\hsize}{%
\includegraphics{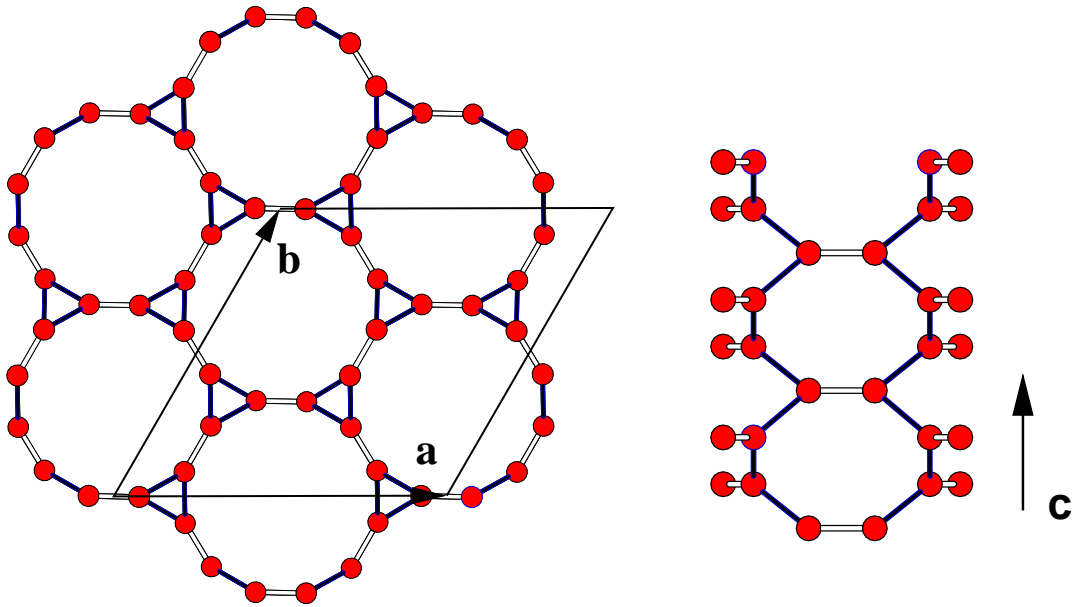}        }}%
\hfill
              }
\caption{ 
An illustration of Cu-sublattice of the dioptase crystal
structure.  The rhombohedral unit-cell contains 18
equivalent Cu atoms arranged in six chains with three atoms
down the $c$ period. 
The inter/intra-chain magnetic coupling with strength $J_1$ and
$J_2$ are indicated by white/black sticks. Left: An $ab$-plane. Not
shown are the Si$_6$O$_{18}$ rings, located inside the 
12-membered Cu-rings. 
The rhombus denotes the in-plane hexagonal unit-cell.
Right: two chiral chains along $c$.
        } 
\label{fig_Cu-lattice}
\end{figure}

We illustrate in fig.~\ref{fig_Cu-lattice} the 
sublattice of the magnetic Cu(II)-ions.
This three-dimensional magnetic lattice is
characterized by only two coupling
constants in between the spin-1/2 Cu(II)-moments.
The magnetic sublattice is characterized
by an antiferromagnetic intra-chain $J_2$, 
which couples the Cu(II)-chains and an antiferromagnetic 
inter-chain coupling $J_1$, leading 
for small $J_1/J_2$ to a AB-type Ne\'el ordered state 
with doubling of the unit-cell along $c$.
Alternatively one might consider the
dioptase magnetic lattice as made up by
in-plane dimers of Cu(II)-ions, with 
an intra-dimer coupling strength of $J_1$
and an inter-dimer coupling along $c$
of $J_2$. For small $J_2/J_1$ a singlet-dimer
state with a spin-gap and no long-range
magnetic order is then realized.

\begin{figure}[t]
\hbox to\hsize{%
\hfill
\rotatebox{0}{\resizebox{!}{0.70\hsize}{%
\includegraphics{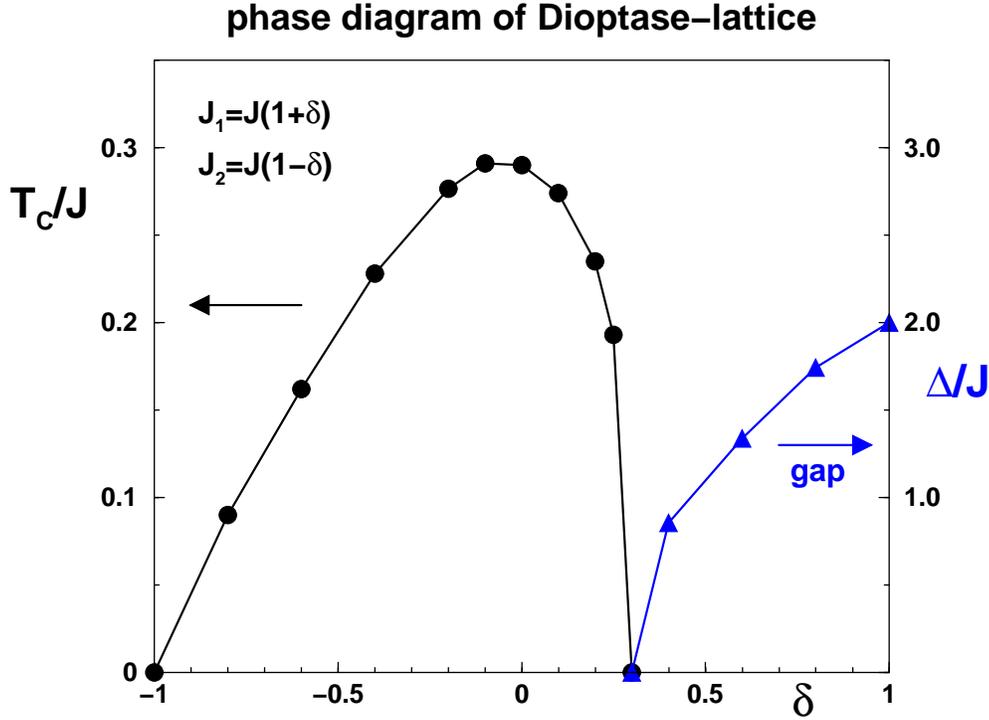}        }}%
\hfill
       }
\caption{Phase diagram of the dioptase magnetic sublattice
         as obtained by Quantum Monte Carlo simulations. 
	 The lines are guides to the eye.
         The magnetic coupling constants are
         $J_{1,2}=J(1\pm\delta)$ for inter/intra-chain
	 couplings $J_1/J_2$. At $\delta_c\approx0.3$ a
	 quantum phase transition occurs. The Ne\'el
	 temperature of the antiferromagnetically ordered
	 phase for $\delta<\delta_c$ is give by the 
	 left $y$-axis. The antiferromagnetic order is of A-B type,
         with a doubling of the unit-cell along $c$.
	 For $\delta>\delta_c$ a gap, given
	 by the right $y$-axis, opens in the magnetic excitation
	 spectrum and the state is a quantum spin-liquid.
        }

\label{fig_phaseDia}
\end{figure}

In fig.~\ref{fig_phaseDia}
we present the phase-diagram of the dioptase magnetic
lattice, which we obtained from Quantum-Monte-Carlo
(QMC) simulations, using the stochastic series 
expansion with worm-updates \cite{san91,san99}.
We used the parameterization $J_{1,2}=J(1\pm\delta)$.

\begin{figure}
\hbox to\hsize{%
  \hfill
\rotatebox{0}{\resizebox{!}{0.55\hsize}{%
\includegraphics{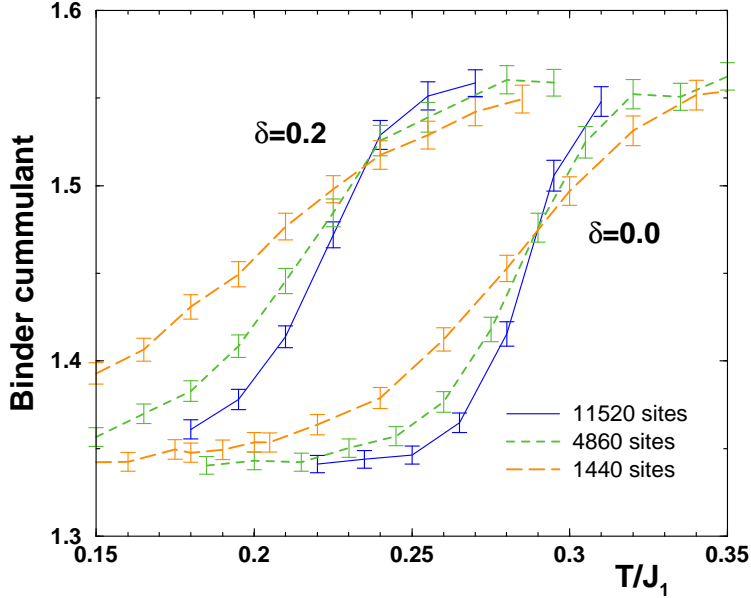}        }}%
    \hfill
    }
\caption{QMC-results for the dimensionless Binder-cumulant
         $\langle m_{AF}^2\rangle / \langle|m_{AF}|\rangle^2$
         for (nnm)-clusters with periodic boundary conditions.
         The lines are guides to the eye, the MC-estimates for
         the statistical errors are given.
         Shown are the results for $n=2$,$m=20$ (1440 sites),
         $n=3$,$m=30$ (4860 sites) and $n=4$,$m=40$ (11520 sites)
         and two value of $\delta$ ($J_{1,2}=J(1\pm\delta)$).
         }
\label{fig_find_Tc}
\end{figure}

In order to determine the phase-diagram, an accurate estimate
of the transition temperature to the ordered state is necessary.
For this purpose we evaluated by QMC one of the
Binder-cumulants \cite{bin97}, namely
$\langle m_{AF}^2\rangle / \langle |m_{AF}|\rangle^2$,
where $m_{AF}$ is the antiferromagnetic-order parameter
(the staggered magnetization). The temperature at which
the cumulants for different finite cluster intersect
provide reliable estimates for the Ne\'el 
temperature \cite{bin97}, see fig.~\ref{fig_find_Tc}.
For the numerical simulations
we used $(n,n,m)$ clusters with periodic boundary conditions,
were $n^2$ and $m$ are the number of unit-cells in the $ab$-plane
and along the $c$-axis respectively. We performed simulations
for $(2,2,20)$, $(3,3,30)$ and $(4,4,40)$ clusters 
containing 1440, 4860 and 11520 Cu(II) sites respectively.

The linear raise of $T_N$ in fig.~\ref{fig_phaseDia}
occurring for small inter-chain couplings $J_1$ 
is a consequence of the quantum-critical nature of the
spin-1/2 Heisenberg chain realized for $J_1=0$.
The magnetic correlation length
$\xi(T)$ diverges as $\xi(T)\sim T^{-1}$ for a Heisenberg-chain at
low-temperature. For small interchain couplings $J_1$
a chain-mean-field approach is valid \cite{sch96}
and the transition occurs when 
$J_2\approx J_1\,\xi(T_N)\sim J_1/T_N$.
Consequently $T_N\sim TJ_1/J_2$ for small $J_1/J_2$.

The critical temperature for the
transition, which is in the 3D-Heisenberg
universality class, is maximal for
$\delta\approx-0.1$ and vanishes at a quantum critical point
$\delta_c\approx0.3$. Long-range magnetism is absent
beyond this point and the ground-state is a quantum
spin-liquid. For $J_2=0$ the dioptase magnetic lattice
decomposes into isolated dimers.

The magnitude of the singlet-triplet gap $\Delta$ 
in the spin-liquid state can be estimated
by a fit of the low-temperature QMC-susceptibility to
$\chi(T)\ \approx\ \left(k_BT / \Delta\right)^{d/2-1}\,
{\rm e}^{-\Delta/(k_B T)}$,
where $d$ is the dimensionality of the triplet-dispersion above
the gap. For an isolated dimer $d=0$, for a spin-ladder
$d=1$~\cite{joh00}. This analysis
would predict $d=3$ for the dioptase magnetic
sublattice, but fits of the QMC-results for $\chi(T)$,
presented in fig.~\ref{fig_phaseDia}, favor $d=0$. 

\begin{figure}
\hbox to\hsize{%
\hfill
\rotatebox{0}{\resizebox{!}{0.70\hsize}{%
\includegraphics{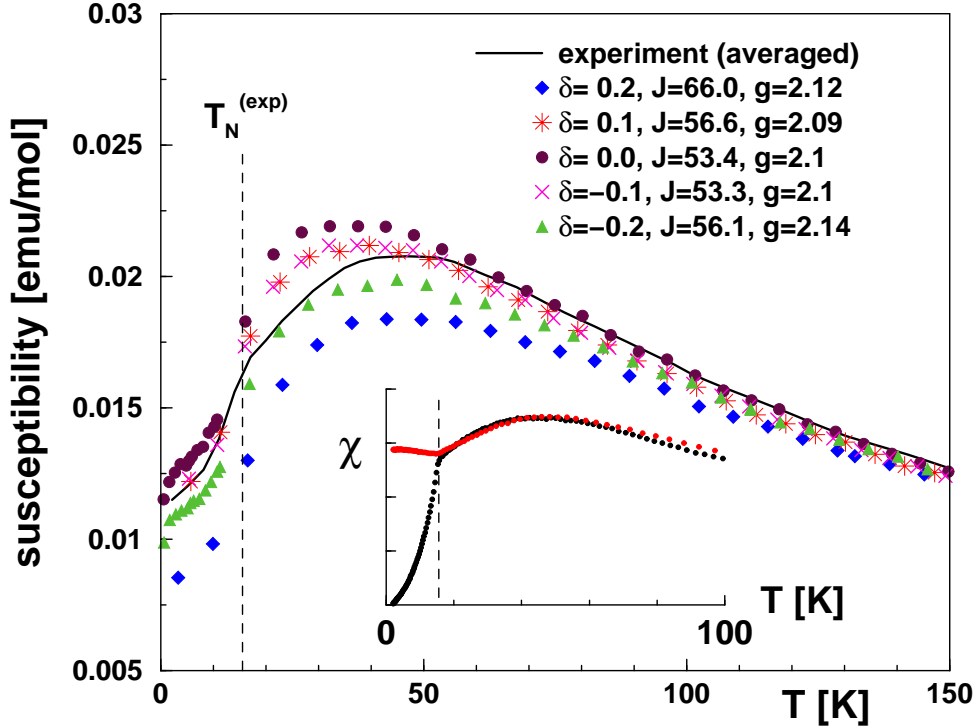}        }}%
\hfill
          }
\caption{QMC-results for the susceptibility (in emu/mol)
         for various $\delta$ in comparison to the
         directional-averaged experimental susceptibility (solid line).
	 Inset: The susceptibility $\chi$ for magnetic fields
	 parallel/orthogonal to the $c$-axis (lower/upper) curve.
	 The vertical dashed lines in the main panel and in the
	 inset indicates the location of the Ne\'el temperature.
        }
\label{fig_susz}
\end{figure}

In fig.~\ref{fig_susz} we present the
susceptibility of green dioptase
(using a crystal from Altyn Tyube, Kazakhstan)
down the He-temperatures measured with a commercial SQUID 
magnetometer (Quantum Design). The data for
magnetic field aligned parallel and perpendicular
to the $c$-axis presented in the Inset of
fig.~\ref{fig_susz} show clearly a transition
to Ne\'el-ordered stated at $T_N^{(exp)}=15.5~{\rm K}$.
The moments are aligned along $c$ for
$T<T_N^{(exp)}$.

The QMC results for the susceptibility are to be compared,
due to spin-rotational invariance, 
with the directional averaged of the experimental
susceptibility, presented in the main panel of
fig.~\ref{fig_susz}.
We have determined the Hamiltonian parameters $J_1=J(1+\delta)$
and $J_2=J(1-\delta)$ appropriate for dioptase in the
following way. For every $\delta<\delta_c$ the overall coupling
constant $J$ was determined by fixing the transition temperature
to the experimental $T_N^{(exp)}=15.5~{\rm K}$. The
spin-susceptibility in experimental units is then

\begin{equation}
\chi^{(exp)}\ =\ 0.375*Z*(g^2/J)*\Lambda_{mm}~,
\label{chi_exp}
\end{equation}

where $Z=3$ is number of Cu$^{2+}$-ions in the primitive
unit-cell. The dimensionless
magnetization-fluctuation is
$\Lambda_{mm} = (J\beta)\left( <m^2>-<m>^2\right)$,
where $m$ is the magnetization.
The $g$-factor was then determined, for
every $\delta<\delta_c$, by adapting the right-hand-side
of eq.~\ref{chi_exp} to the experimental susceptibility at
high temperatures.
The results are shown in
fig.~\ref{fig_susz} together with the optimal values for $J$ and $g$.
We see that the optimal value $g\approx2.1$ for the $g$-factor is
relatively independent of $\delta$. 


We find two possible values for the ratio of
the two-coupling (antiferromagnetic)
constants $J_1$ and $J_2$ namely $\delta=0.1$ and
$\delta=-0.1$ which fit the experimental data equally well.
Note that $\delta=-0.2$ does not agree well for $T<T_C^{(exp)}$.
We attribute the residual discrepancies in between the theory and
the experimental data to residual interactions, in addition
to $J_1$ and $J_2$

It has been suggested previously~\cite{wit93} that the
in-chain coupling $J_2$ might actually be ferromagnetic. We have
studied therefore also the case for negative $J_2$ and found a
quantum-phase-transition to a state with alternating
ferromagnetic chains for $J_2\approx-0.7\,J_1$. We have performed
the corresponding analysis to the one shown in fig.~\ref{fig_susz}
for the the case of ferromagnetic $J_2$. 
We found very large deviations in between
experiment and theory in this case, due to the fact that the
susceptibility of ferromagnetic chains diverges for $T\to0$.

\begin{figure}
\hbox to\hsize{%
\hfill
\rotatebox{0}{\resizebox{!}{0.70\hsize}{%
\includegraphics{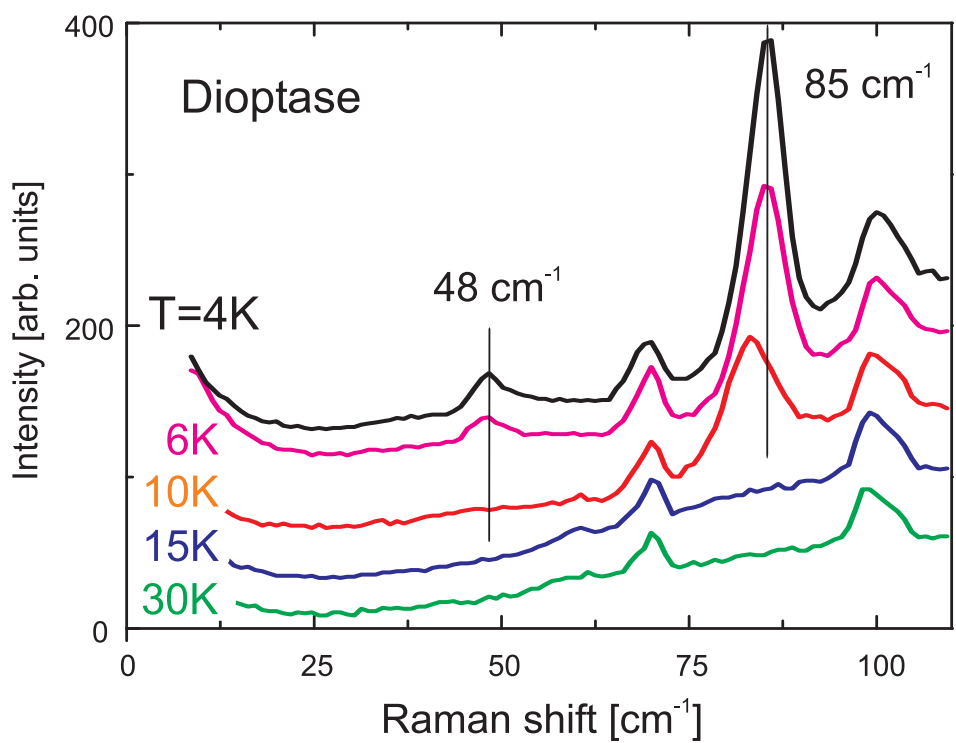}        }}%
\hfill
          }
\caption{Low energy Raman spectrum of dioptase in xx-polarization. 
The modes at 48 and 85~cm$^{-1}$ ($\equiv$~69 and 122~K) show a 
strong increase of intensity for $\rm T<T_{N}=15.5\,K$ and
correspond to one- and two-magnon processes. The temperature 
independent modes at 70 and 100~cm$^{-1}$ are phonons. }
\label{raman}
\label{fig_raman}
\end{figure}

To settle the ambiguity concerning the $\delta$ 
parameter we investigated the magnetic Raman spectrum 
of dioptase as a function of temperature,
as shown in fig.~\ref{fig_raman}.  
The Raman scattering experiments were
performed in quasi-backscattering geometry with a triple grating
optical spectrometer (DILOR XY) with the $\lambda$~=~514~nm laser line.
Two modes at 48 and 85~cm$^{-1}$
($\equiv$~69 and 122~K) are magnetic as they
exhibit a temperature dependence related to
the transition. They show no anisotropy concerning 
the scattering selection rules. The excitation
energies 69~K and 122~K correspond, for $\delta=+0.1$, 
to one and two inter-chain dimer excitation energy 
$J_1=J(1+\delta)$, as expected for one- and 
two-magnon scattering processes. The lineshape of
the magnetic two-magnon 122~K mode is very unusual,
it is symmetric and not substantially broaded by
either magnon-magnon scattering or density-of-states
effects, in contrast to usual two-magnon scattering
in normal 3D antiferromagnets \cite{fle70}. This
behavior indicates a very small 
dispersion of the underlying magnon branch.
We consequently conclude that dioptase is relatively close
to a quantum-critical point.

In conclusion we have presented a novel magnetic lattice
structure, the dioptase magnetic lattice, which allows
for a quantum-phase transition. This lattice is realized
in green dioptase $\rm Cu_6Si_6O_{18}\cdot 6H_2O$ and
in the recently synthesized isostructural germanate
$\rm Cu_6Ge_6O_{18}\cdot 6H_2O$ \cite{bra97,bae01},
a promising candidate to study further aspect of the
phase diagram presented in detail fig.~\ref{fig_phaseDia}.


We acknowledge fruitful discussions with Matthias
Troyer on the stochastic series expansion and Felicien Capraro
for data analysis.


\end{document}